\documentclass[runningheads]{llncs}
\usepackage{graphicx}
\begin{document}
\title{Testing the robustness of attribution methods for convolutional neural networks in MRI-based Alzheimer's disease classification}
\titlerunning{Attribution robustness in AD classification}
%
\author{Fabian Eitel\inst{1} \and
Kerstin Ritter\inst{1} for the Alzheimer's Disease Neuroimaging Initiative (ADNI)
}
\authorrunning{F. Eitel et al.}
%
\institute{
Charit\'e -- Universit\"atsmedizin Berlin, corporate member of Freie
Universit\"at Berlin, Humboldt-Universit\"at zu Berlin, and Berlin Institute of
Health (BIH); Department of Psychiatry and Psychotherapy; Berlin Center for Advanced Neuroimaging, Bernstein Center for Computational Neuroscience; 10117 Berlin, Germany.
}
\maketitle              
\begin{abstract}
Attribution methods are an easy to use tool for investigating and validating machine learning models. Multiple methods have been suggested in the literature and it is not yet clear which method is most suitable for a given task. In this study, we tested the robustness of four attribution methods, namely  gradient*input, guided backpropagation, layer-wise relevance propagation and occlusion, for the task of Alzheimer's disease classification. We have repeatedly trained a convolutional neural network (CNN) with identical training settings in order to separate structural MRI data of patients with Alzheimer's disease and healthy controls. Afterwards, we produced attribution maps for each subject in the test data and quantitatively compared them across models and attribution methods. We show that visual comparison is not sufficient and that some widely used attribution methods produce highly inconsistent outcomes.

\keywords{machine learning  \and convolutional neural networks \and MRI \and explainability \and robustness \and attribution methods \and Alzheimer's disease}
\end{abstract}
\section{Introduction}
As machine learning becomes more and more abundant in medical imaging, it is necessary to validate its efficacy with the same standards as other techniques. On magnetic resonance imaging (MRI) data, several studies have reported classification accuracies above 90\% when using machine learning to detect neurological and psychiatric diseases (for a review, see \cite{Vieira2017}). While these results seem promising at first, an in-depth investigation of those results both in terms of generalizability as well as medical validity is necessary before they can enter clinical practice. Medical validity can be examined by using attribution methods such as saliency analysis. Specifically, the decision of a machine learning algorithm can be visualized as a heatmap in the image space, in which the contribution of each voxel is determined. To identify the relevance of specific brain areas, quantitative and qualitative analyses on the heatmaps can be performed. Models that shift importance to areas which are well known to be clinically relevant in specific diseases might be more suitable for clinical practice in comparison with models that scatter relevance across the entire image or to seemingly random brain areas. While it might not be necessary to understand the exact workings of a model, similar to many drugs used in clinical practice, the causal mechanism of a model should have at least a minimal coherence with the causal reasoning of a clinical expert and should be interpretable by the expert.

In neuroimaging studies, where sample sizes are often extremely limited, specific attention needs to be given to robustness. Small sample sizes can cause model training to be rather fluctuating and varying between different runs. One can avoid ``cherry-picking'' of final results easily by identically repeating training procedures and reporting average scores. In doing so, the question arises whether attribution methods suffer from similar variances. 
In the present study, we therefore propose to evaluate the \textit{robustness} of attribution methods.
Specifically, we investigate whether multiple heatmap methods are coherent in their results over identical training repetitions with a variety of measures. For this purpose, we trained a convolutional neural network (CNN) several times to separate structural MRI data of patients with Alzheimer's disease (AD) and healthy controls. For each subject in the test data, we then produced heatmaps using four widely used attribution methods, namely gradient*input, guided backpropagation, layer-wise relevance propagation (LRP) and occlusion. All those methods have been applied in MRI-based AD classification before \cite{korolev2017,rieke2018visualizing,boehle2019visualizing}.  
As it was noted in \cite{Sundararajan2017axiomatic} specific criteria are needed in order to avoid artifacts from the data, the model or the explanation method in order to empirically compare them. Here we point out the issue of artifacts from model training and present a framework to investigate them.

\section{Related Work}
Different criteria for evaluating visualization methods have been proposed in the literature, including sensitivity, implementation invariance, completeness and linearity \cite{Sundararajan2017axiomatic}, selectivity \cite{Bach2015LRP}, conservation and positivity \cite{MONTAVON2017taylordecomposition} as well as continuity \cite{MONTAVON2018interpreting}. Additionally, \cite{adebayo2018sanitychecks} has introduced two sanity checks of visualization methods based on network and data permutation tests. Only \cite{alvarez2018robustness} has investigated robustness so far. We differ from \cite{alvarez2018robustness} by repeating the training cycle and comparing the outcomes without any perturbation. 

In neuroimaging, only a few studies have compared attribution methods. \cite{rieke2018visualizing} has given an overview of four different attribution methods for MRI-based Alzheimer's disease classification and introduced a modified version of occlusion, in which brain areas according to an atlas are occluded. For the same task, \cite{boehle2019visualizing} has presented an in-depth analysis together with multiple metrics for evaluating attribution methods based on LRP and guided backpropagation as a baseline method. In \cite{eitel2019uncovering}, it has been shown that LRP and gradient*input led to almost identical results for MRI-based multiple sclerosis detection.

\section{Methods}
The dataset used in this study is part of the Alzheimer's Disease Neuroimaging Initiative\footnote{http://adni.loni.usc.edu/} (ADNI) cohort. Specifically, we have collected 969 T1-weighted MPRAGE sequences from 344 participants (193 AD patients and 151 healthy controls) of up to three time-points. The full-sized 1mm isotropic images were non-linearly registered using the ANTs framework to the ICBM152 (2009c) atlas. We have split the dataset patient-wise by sampling 30 participants from each class into a test set and 18 participants from each class into a validation set. All available time-points were then used to increase the total sample size. Additionally, the data was augmented by flipping along the sagittal axis with a probability of 50\% and translated along the coronal axis between -2 and 2 voxels.

The 3D-CNN used to separate AD patients and healthy controls  consists of 4 blocks of Conv-BatchNorm-ReLU-MaxPool followed by two fully-connected layers, the first being activated by a ReLU as well. A dropout of 0.4 was applied before each fully-connected layer. All convolutional layers use 3x3x3 filters, with 8, 16, 31, 64 filters from bottom to top layers. Max pooling uses a pooling size of 2, 3, 2, 3 voxels respectively. We used the ADAM optimizer 
with an initial learning rate of 0.0001 and a weight decay of 0.0001. Furthermore, early stopping with a patience of 8 epochs was employed. The training was repeated for 10 times to create 10 identically trained models, albeit each randomly initialized. Note that mini-batch ordering was not fixed between different runs.

For each trained model and each subject in the test set, we produced heatmaps using the following attribution methods: 

{\bfseries Gradient * input \cite{shrikumar2017learning}} multiplies the gradient, which has been backpropagated into the input space, with the original input. It is an adaption of saliency maps \cite{simonyan2013deep} and increases the sharpness of the resulting heatmaps. 

{\bfseries Guided backpropagation \cite{springenberg2015guidedbackprop}} modifies the backpropagation in ReLU layers by only passing positive gradients. Since the backpropagation ignores features in which the ReLU activation is zero, guided backpropagation requires both the gradient and the activation to be non-zero.

{\bfseries Layer-wise relevance propagation (LRP) \cite{Bach2015LRP}} backpropagates the classification score instead of the gradient and multiplies it with the normalized activation for each neuron. LRP conserves the relevance under certain conditions such that the sum of relevance for all neurons does not change between layers.

{\bfseries Occlusion \cite{zeiler2014visualizing}} is, unlike the other presented methods, not based on backpropagation. In occlusion, the attribution is computed by the change in the output score, when some part of the input example is ``occluded'' (i.e. set to zero). Here, we occlude a volumetric patch which is shifted over the entire MRI volume. Although the occlusion method results commonly in much coarser heatmaps (depending on the size of the patch), we included this method because it has been used several times in MRI-based AD classification \cite{Esmaeilzadeh2018,korolev2017,Liu2018,rieke2018visualizing}.


Besides comparing the attribution maps directly, we have carried out atlas-based comparisons using the Neuromorphometrics atlas \cite{bakker2015scalable} in which left and right hemisphere regions have been combined. We computed the attribution within each region based on three metrics: {\bfseries sum} as the sum of absolute values in each region, {\bfseries density} as the regional mean, i.e. the sum normalized by the size of the respective region, as well as {\bfseries gain} as the ratio between the sum for patients divided by the sum for healthy controls. The latter was defined in \cite{boehle2019visualizing} arguing that healthy controls typically also receive positive relevance. By normalizing each region to a control average those regions which exhibit strong differences between controls and patients are highlighted.
When sorting brain areas by these metrics, we therefore obtain three rankings for each repetition and method. We then compare the intersection between the top 10 regions of each repetition in order to see whether repeated runs highlight similar regions.

\section{Results}
The balanced accuracy of all 10 training runs on the test set is on average 86.74\% with a considerable range of 83.06\% to 90.12\% between runs. 

\begin{figure}
\begin{center}
\includegraphics[width=\textwidth]{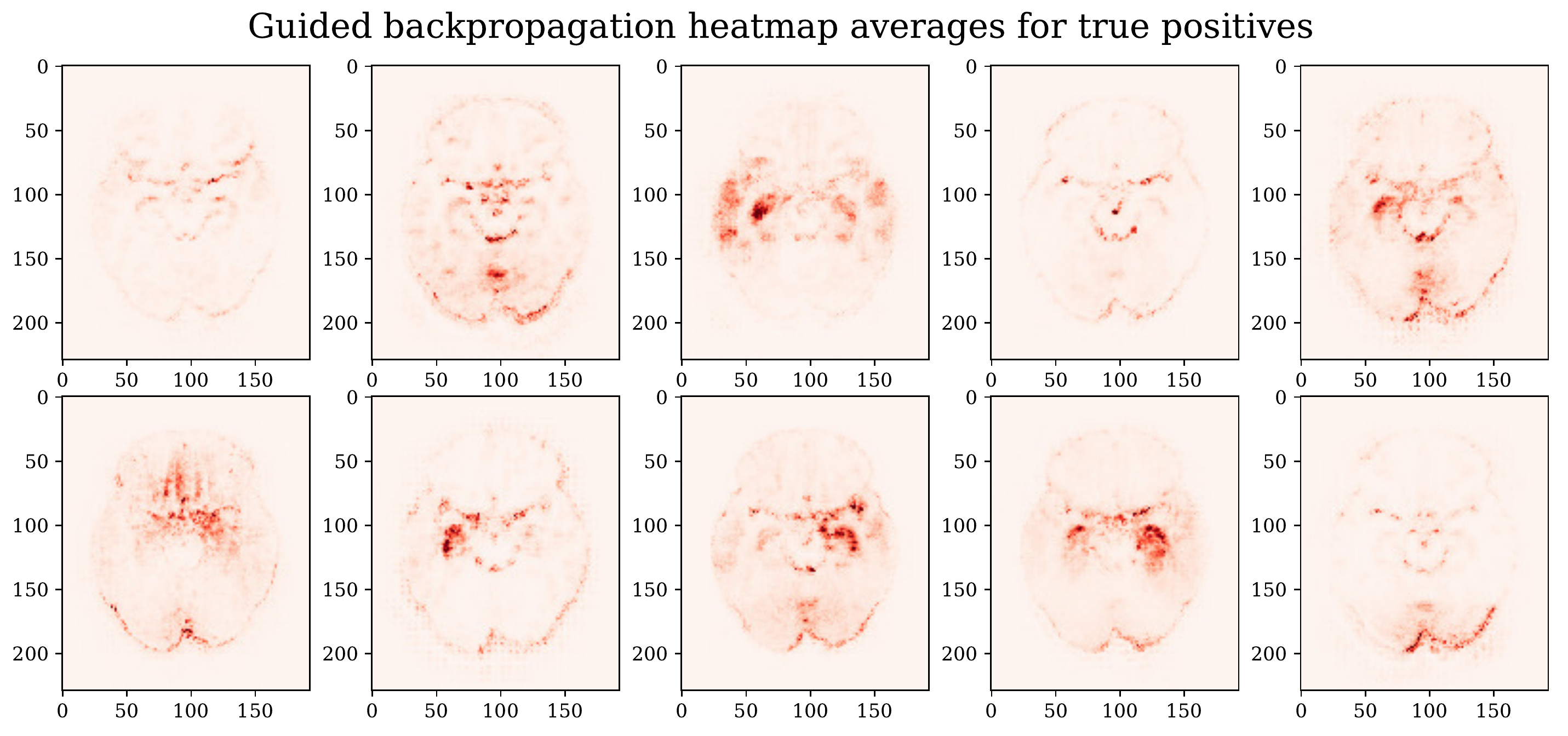}
\caption{Different heatmap outcomes (averaged over all true positives) for guided backpropagation and each of the 10 trained models.} 
\label{fig:GB_only}
\end{center}
\end{figure}

Figure \ref{fig:GB_only} shows the guided backpropagation attribution maps, averaged over all true positives for each of the 10 training runs. Solely by visually inspecting them, one can see clear differences between the various runs. While some heatmaps seem to highlight the hippocampus (top row middle, bottom row 2\textsuperscript{nd} to 4\textsuperscript{th}) others do not (top row 1\textsuperscript{st} and 4\textsuperscript{th}, bottom row 5\textsuperscript{th}). In almost all heatmaps the edges of the brain are given attribution and in some a large amounts are given to the cerebellum (bottom row 1\textsuperscript{st}, 2\textsuperscript{nd} and 5\textsuperscript{th}). The heatmaps from the other methods exhibit similar variances.


\begin{figure}
\begin{center}
\includegraphics[width=0.95\textwidth]{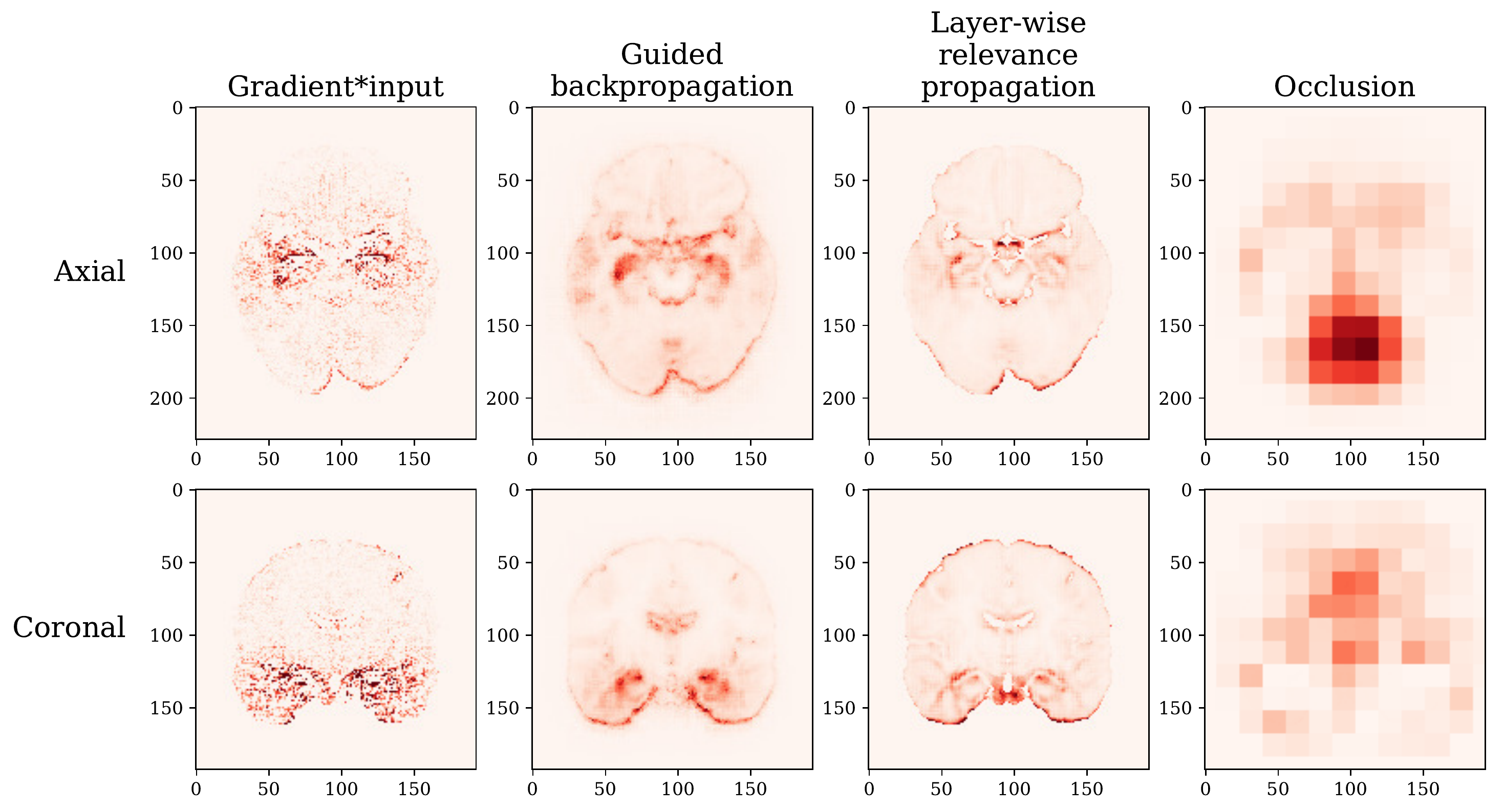}
\caption{Averages over all true positive attribution maps and all 10 runs for each attribution method. } \label{fig:compare}
\end{center}
\end{figure}

In Figure \ref{fig:compare}, we compare the four different attribution methods. Occlusion clearly stands out by producing a much coarser attribution map than the other methods. This is due to the fact that the size of the patch which is being occluded, scales inversely with the run-time. Running the occlusion method for a 3D image with a patch size of 1x1x1, in order to match the sharpness of the other methods, would be computationally unfeasible. One can also note that gradient*input seems to produce the least continuous regions. With exception of occlusion, all methods seem to attribute importance to the hippocampus.

An attribution method would be expected to produce similar heatmaps when the model training is repeated identically. Table \ref{tabl:norms} shows the L2-norms for each attribution method, between all of its average attribution maps. LRP and guided backpropagation have the smallest L2-norms between their average heatmaps. Occlusion has L2-norms by a magnitude larger than the other methods, which might be due to the limited sharpness. 
Average heatmaps have been scaled by their maximum value to produce comparable results.
\begin{table}
\begin{center}
\caption{L2-norm between average attribution maps of all different runs for true positive and true negative predictions.}\label{tabl:norms}
\begin{tabular}{|l|c|c|}
\hline
 Method & True positives & True negatives\\
\hline
 Gradient * input & 3102 & 3145\\
 Guided backpropagation & 2930 & \textbf{1992} \\
 LRP & \textbf{2241} & 2196\\
 Occlusion & 25553 & 30774\\
\hline
\end{tabular}
\end{center}
\end{table}

\begin{figure}
\includegraphics[width=\textwidth]{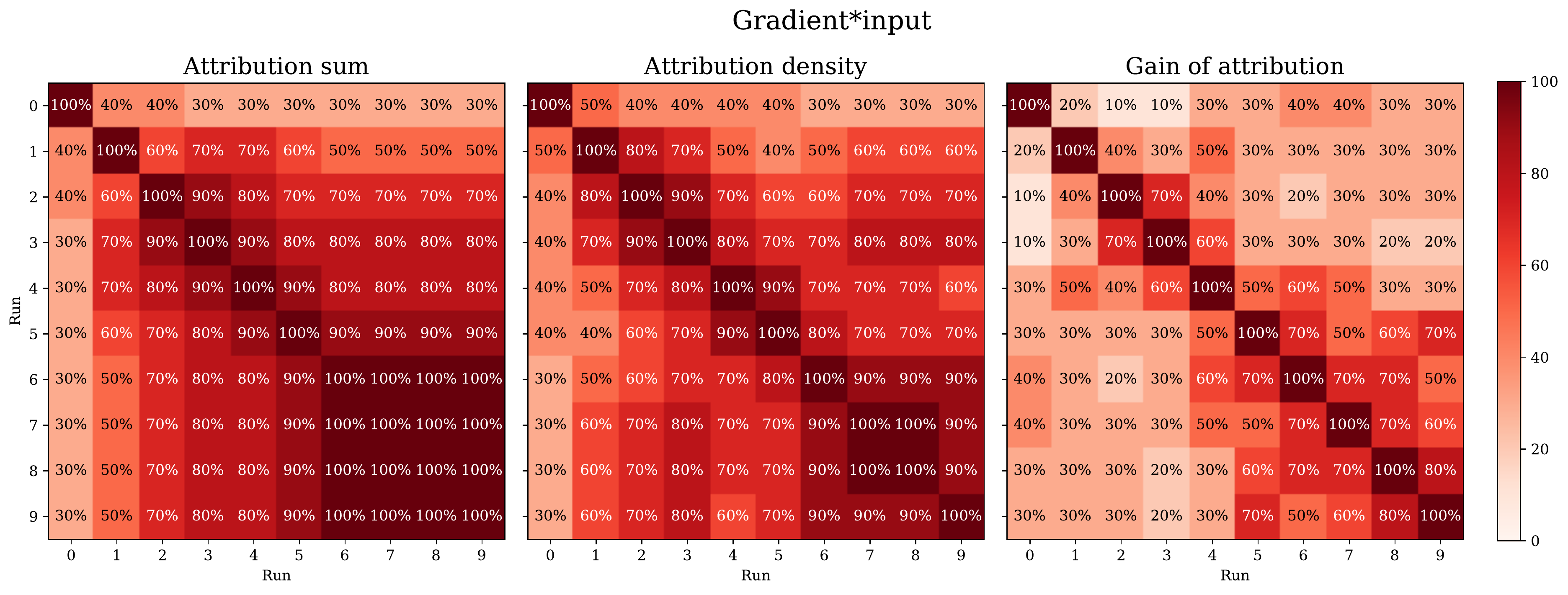}
\includegraphics[width=\textwidth]{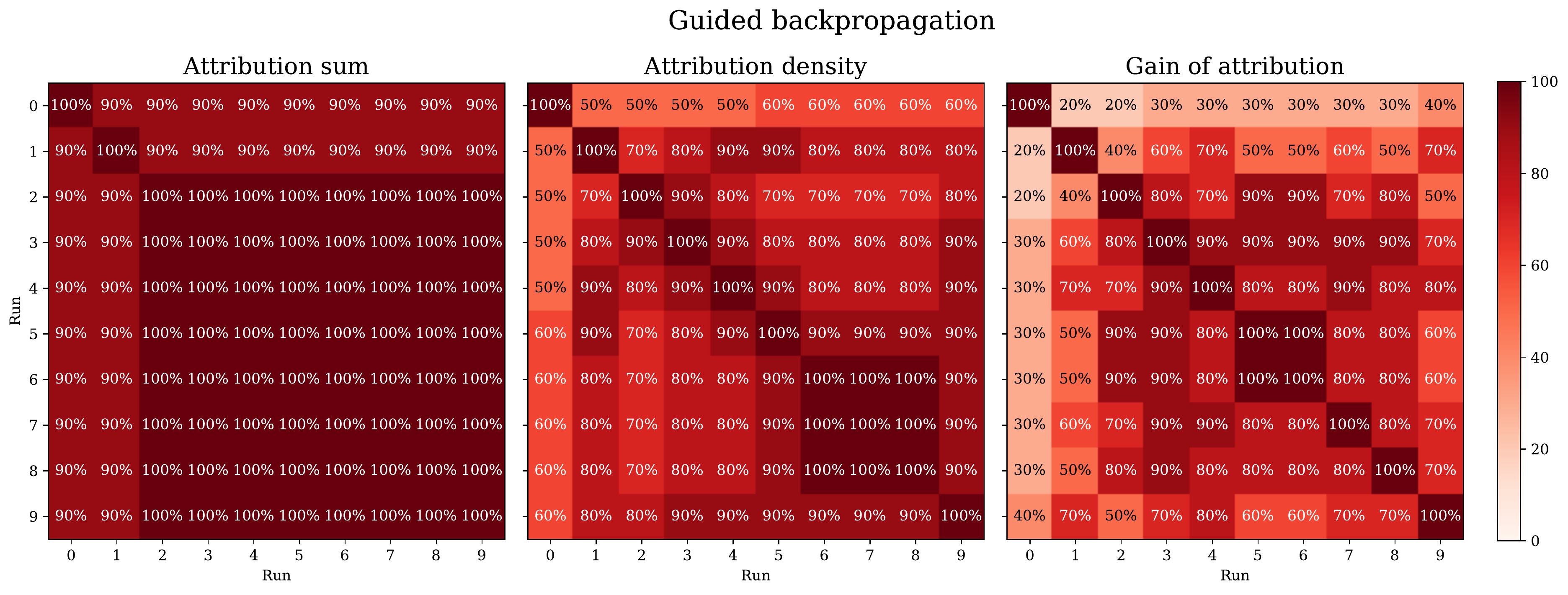}
\includegraphics[width=\textwidth]{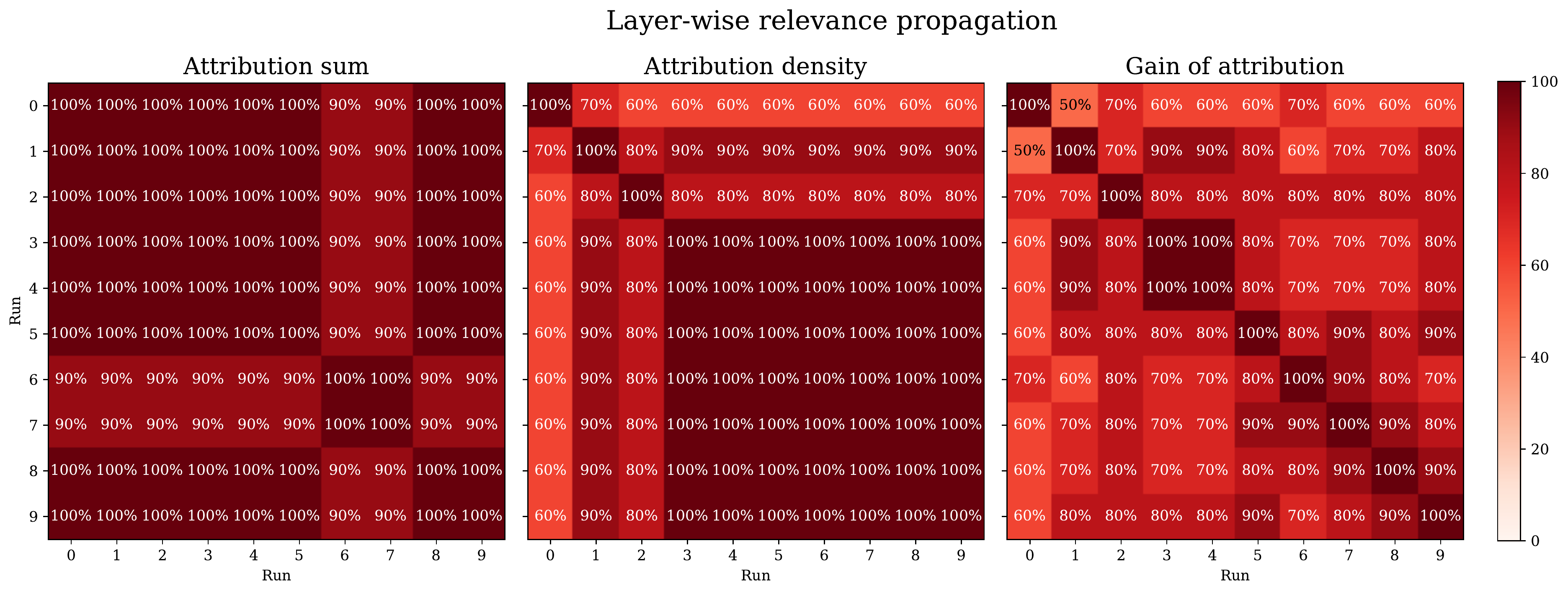}
\includegraphics[width=\textwidth]{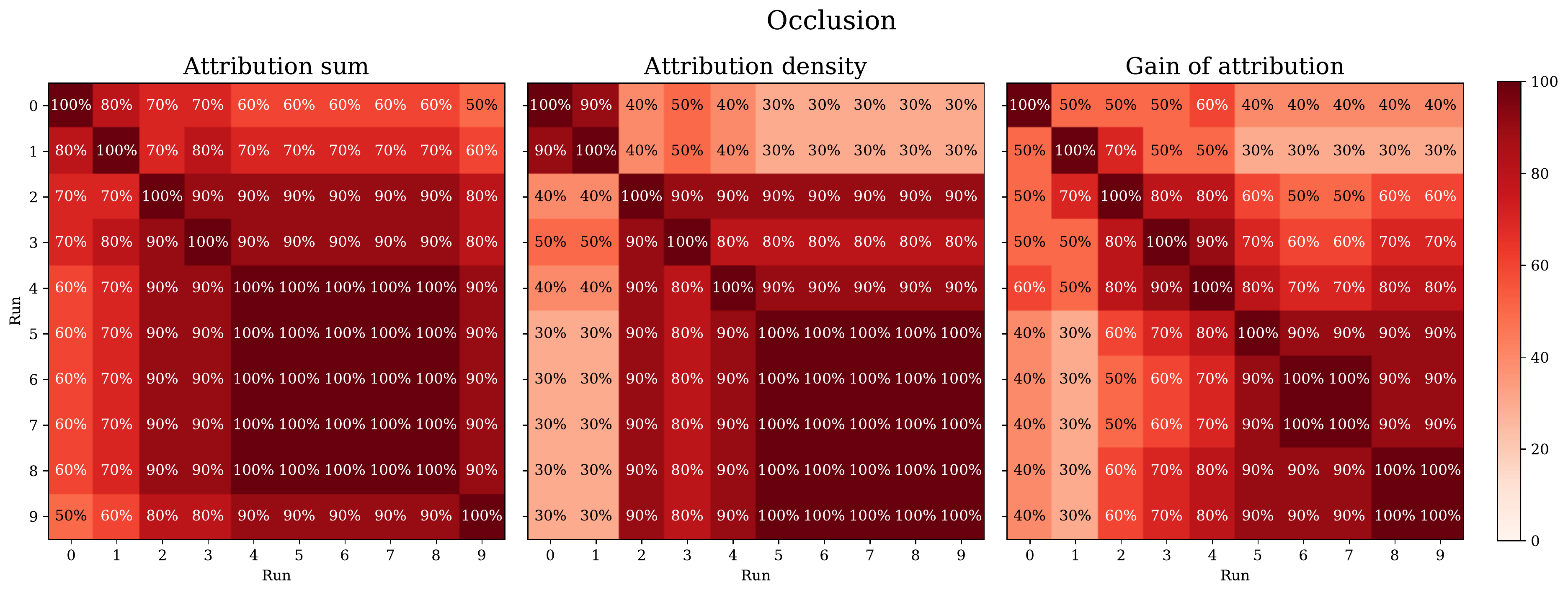}
\caption{Intersection between the 10 regions with the highest attribution according to the total sum, the size-normalized density and the control-normalized gain.} \label{fig:intersections}
\end{figure}

When dividing the attributions into brain regions, large regions such as cerebral white matter and the cerebellum receive most attribution. Normalized by region size, the basal forebrain, 4\textsuperscript{th} ventricle, hippocampus and amygdala become highlighted. Standardizing by attributions of healthy controls leads to rather inconsistent orderings.
In Figure \ref{fig:intersections}, we show how much the top 10 brain regions, in terms of attribution, intersect with each other over repetitions. An intersection of 100\% means that the regions between those two runs contain the same regions in their top 10, ignoring the order within those 10. In Table \ref{tabl:top10averages}, we averaged the intersections separately for each attribution method and each metric. All methods have their highest intersection in terms of region-wise sum, guided backpropgation and LRP seem to reproduce the same regions almost perfectly. Even though occlusion seemed to perform poorly in the other measures it has a consistency higher than gradient*input. All methods perform worst in terms of gain of relevance which might be due to the scarcity within healthy control attribution maps as discussed in \cite{boehle2019visualizing}.

\begin{table}
\begin{center}
\caption{Averages of top 10 region coherence.}\label{tabl:top10averages}
\begin{tabular}{|l|c|c|c|}
\hline
 Method & Attribution sum & Attribution density & Gain of attribution \\
\hline
 Gradient * input & 72.60 \% & 69.00 \% & 46.40 \% \\
 Guided backpropagation & 96.60 \% & 80.60 \% & 68.60 \% \\
 LRP & 96.80 \% & 88.40 \% & 78.00 \% \\
 Occlusion & 84.60 \% & 74.20 \% & 67.80 \% \\
\hline
\end{tabular}
\end{center}
\end{table}

\section{Discussion}
In this study, we have shown that attribution methods differ in robustness with respect to repeated model training. In particular, we found that LRP and guided backpropagation produce the most coherent attribution maps, both in terms of distance between attribution maps as well as in terms of order of attribution awarded to individual regions. We also confirm that solely visually judging heatmaps is a deficient criteria as pointed out by \cite{adebayo2018sanitychecks}. Especially in medical imaging, it is important to acknowledge that the small sample sizes available lead to variances in the output. These variances make it hard to compare and to replicate outcomes of individual studies. Even though reporting metrics averaged over repeated training runs is an effective tool to reduce the variances, it is rarely used in the community. Here, we have extended the repetition to attribution methods and shown similar variances. Even though these variances likely stem from the different local minima each training run ended up in, attribution methods which cease the variances and report similar outcomes are highly preferable. In conclusion, we think that domain specific metrics, as suggested in this study, are essential for identifying suitable attribution methods.



%
%

%
%
%
 \bibliographystyle{splncs04}
 \bibliography{references.bib}
%
\end{document}